\documentclass{elsarticle}
\usepackage[utf8]{inputenc}
\usepackage{natbib}
\usepackage{textgreek}
\usepackage{subfigure}
\usepackage[a4paper, total={7in, 10in}]{geometry}
\usepackage[table,xcdraw]{xcolor}
\title{Adapting U-Net for linear elastic stress estimation in polycrystal Zr microstructures}
\author{J. D. Langcaster, D. S. Balint, M. R. Wenman }
\date{October 2022}

\usepackage{amsmath}
\usepackage{amssymb}
\usepackage[breaklinks]{hyperref}
\usepackage{dsfont}
\usepackage{graphicx}
\usepackage{multirow}
\usepackage{pdflscape}

\newcommand{\MSE}{\mathrm{MSE}}

\begin{document}

\maketitle

\section{Abstract}
A variant of the U-Net convolutional neural network architecture is proposed to estimate linear elastic compatibility stresses in α-Zr (hcp) polycrystalline grain structures. Training data was generated using VGrain software with a regularity $\alpha$ of 0.73 and uniform random orientation for the grain structures and ABAQUS to evaluate the stress fields using the finite element method. The initial dataset contains 200 samples with 20 held from training for validation. The network gives speedups of around 200x to 6000x using a CPU or GPU, with significant memory savings, compared to finite element analysis with a modest reduction in accuracy of up to 10\%. Network performance is not correlated with grain structure regularity or texture, showing generalisation of the network beyond the training set to arbitrary Zr crystal structures. Performance when trained with 200 and 400 samples was measured, finding an improvement in accuracy of approximately 10\% when the size of the dataset was doubled.

\section{Keywords}
Microstructure, Machine Learning, Finite Element Analysis, Zirconium, Nuclear Materials

\section{Introduction}
Internal microstructural type II stresses \cite{Withers2001ResidualTechniques} in Zr-based nuclear fuel cladding in water-cooled reactors has a significant impact on the onset and progress of hydrogen diffusion and hydride precipitation that may lead to cladding failure. In particular, grain compatibility stresses in Zr-based alloys influence the diffusion of hydrogen within the matrix, leading to the concentration of hydrogen in regions of tensile hydrostatic stresses and allowing the formation of hydrides, causing either embrittlement or delayed hydride cracking (DHC) \cite{Liu2022CharacterisationZircaloy-4} . As such, there is interest in the modelling of these compatibility stresses to feed into microstructurally sensitive models of hydride precipitation. Currently compatibility stresses can be estimated using finite element (FE) analysis, though such simulations can only feasibly simulate small domains on the order of microns with a few hundred grains. This means that larger scale statistical analysis based upon macroscopic measures of microstructure are impractical even if they are desired for industrial applications. Recent headway by Patel et al. \cite{Patel2021AAlloys} has been made in the microstructural modelling of hydride precipitation in a 2D cross section of the cladding, which incorporates some microstructural features such as grains, crystallographic microhydride precipitation and notch defect stress fields, and can be executed in a few seconds. However, this model lacks awareness of grain compatibility stresses that generally require a full numerical solution, such as by FE methods, resulting in slow execution, which the model was designed to avoid. It is within these restrictions that a machine learning based surrogate can provide an alternative, which is investigated in this paper.

In the last decade, machine learning based methods have been shown to be capable of tackling a wide array of real world problems, such as image domain transfer e.g. segmentation of cell micrographs \cite{Ronneberger2015U-net:Segmentation} or diagnosis of COVID-19 using chest x-rays \cite{Rahaman2020IdentificationApproaches}. Very recently, generalised network architectures have been shown to be effective at emulating physical simulations across a variety of domains with very limited data \cite{Kasim2022BuildingSearch}, displaying the effectiveness of machine learning techniques for physical problems. Once trained, these algorithms can be orders of magnitude faster than the simulations they are based on, which can save significant CPU time if the model can generalise beyond the training data.

This paper proposes the use of a convolutional neural network based upon U-Net \cite{Ronneberger2015U-net:Segmentation} to accelerate the prediction of the maximum principal stress within an α-Zr hcp grain structure given only a map of grain orientation. Using a machine learning algorithm can bypass the need for an expensive FE or equivalent solver beyond generating training data. Provided the trained algorithm can be generalised beyond the grain structures in the training data then significant speed-ups in calculation can be realised with a small sacrifice in accuracy. A trained network may then be incorporated into a model, such as that by Patel et al. \cite{Patel2021AAlloys} to approximate these stresses to inform the hydride precipitation model without sacrificing the execution speed of the model.

U-Net \cite{Ronneberger2015U-net:Segmentation} was originally designed for the segmentation of cancer cell micrographs as an image to image mapping solver. It utilises a convolutional encoder-decoder layout, in which all network layers are parameterised convolutions with downsampling and upsampling steps in the encoder and decoder sections respectively. In addition, the network also feeds outputs from earlier in the network into later layers, which helps the network restore smaller scale details in the output which may be missed during the upsampling step, resulting in sharper output images. U-Net has been extended to several other fields beyond image segmentation, such as solving for the temperature field given a map of sources \cite{Zhao2023Physics-informedData}. U-Net has also been used as a mesh super-resolution method for stress fields \cite{Xu2022SuperMeshing:Simulation}, where the output of a low resolution FE simulation is refined by a U-Net to match the output of a high resolution FE simulation without the required computation time and memory requirements.

Traditional simulations rely on discretising the partial differential equations that describe the physical behaviour in question. By discretising, the PDEs are transformed into a system of finitely many algerbraic equations, amenable to solution by a computer. The finite element method (FEM) is a commonly used discretisation and solution method with significant flexibility in how it divides the input domain into finite elements. Overall, the solver of a PDE system can be expressed as an operator $G$ which maps the input, in this case a map of the orientation or more generally the elasticity tensor of the domain $\phi$ to the desired output, in this case the stress state $\mathbf{\sigma}$, such that $G\left(\phi\right) = \mathbf{\sigma}$. Numerical solvers are an explicitly programmed approximate operator based upon discretisation of the PDE system while the goal of this paper is to propose a neural network that can be trained to approximate the operator $G$.

Neural networks can vary in size from a handful of parameters for simple classification systems up to hundreds of billions of parameters in the most complex language models \cite{Brown2020LanguageLearners}, but typical convolutional neural networks for parsing spatial data usually have a few million parameters. The basic unit of a neural network, the eponymous neuron, is a simple linear combination of its input vector, $\vec{x}^{n-1}$, multiplied by a weight matrix, $\mathbf{W}^n$ and added to a bias vector, $\vec{b}^n$. This linear combination is then passed through a non-linear activation function, $\sigma^n$, to produce the output vector, $\vec{x}^{n}$, where $n$ is the layer number.
\begin{equation}
    \vec{x}^{n} = \sigma^n \left(\mathbf{W}^n \cdot \vec{x}^{n-1} + \vec{b}^n \right)
\end{equation}
The activation function must be non-linear in order for the network to learn non-linear relationships in the data. However the degree of non-linearity depends ultimately on the specific data required for the model. While sigmoid functions were often used in earlier neural networks for all layers \cite{Rumelhart1986LearningPropagation}, recently the rectified linear unit (ReLU) \cite{Glorot2011DeepNetworks} has been favoured due to both its performance in fitting the data \cite{Hanin2017ApproximatingWidth} and computational efficiency. A neural network is then the composition of many of these layers, resulting in a highly non-linear function, which can approximate any arbitrary function, as outlined in the universal approximation theorem \cite{Hanin2017ApproximatingWidth} \cite{Cybenko1989ApproximationFunction} . The weight matrices and bias vectors are collectively the parameters of the network that allows the network to fit the specific problem at hand. The overarching network architecture may be encoded collectively as the network's hyperparameters, the optimisation of which is the subject of active research \cite{Kasim2022BuildingSearch} \cite{Gong2019AutoGAN:Networks}, although currently most models are tuned by hand for the specific problem.

The problem of finding suitable parameters for a specific task can be formally encoded as finding the parameters, $\theta$, of a function, $f$, that maps from the input set, $X$, to the output set, $Y$, by means of minimisation of a cost function, $C$, which is a metric between two points in $Y$.
\begin{equation}
    \theta^* = \min_\theta C\left(f_\theta\left(X\right), Y\right)
\end{equation}
The parameters are typically found using a variant of the gradient descent method via the backpropagation of gradients through to earlier parts of the network \cite{Rumelhart1986LearningPropagation}, taking a small step down the gradient in the parameter space.
\begin{equation}
    \theta^{t+1} = \theta^t - \lambda \nabla_\theta C\left(f_\theta(X), Y\right)
\end{equation}
The size of the step is determined by the learning rate, $\lambda$, which may be set at training initiation or changed during runtime to fit the problem.

The cost function has a significant effect on convergence of the network since it determines the cost landscape in which the parameters must be optimised, which can be quite particularly rough for some network architectures \cite{Li2018VisualizingNets}. For regression type problems, the mean squared error (MSE) provides a stable measure of difference with the added benefit that the cost is approximately quadratic around local minima, such that the gradient naturally tends towards zero at the minima. As a result, learning rate does not need to be adjusted to allow convergence. Another commonly used cost function is the Kullback-Leibler divergence \cite{Kullback1951OnSufficiency}, which measures the relative informational entropy between 2 distributions, making it well suited to dealing with Bayesian Learning algorithms \cite{Buntine1991BayesianBack-Propagation}, which encode weights as probability distributions rather than single numbers. 

As the datasets used for training are often much larger than available GPU memory (VRAM), especially when tracking parameter gradients, the dataset is often broken into smaller batches to perform each step, which introduces some noise into the gradient but allows for faster step times and lower memory requirements.

Connectivity of neurons between layers has significant influence on the characteristics of the network and which tasks the network is suited to tackling. Most networks in early research had each neuron in a layer connected to all neurons in the subsequent layer, known as fully connected (FC) layers. This comes with several disadvantages when applying the network to image or volumetric type data. First, the number of parameters in the network quickly balloons with the number of neurons in each layer, which becomes a significant problem when the input data is, for example, a 1000x1000 pixel image, requiring a million neurons in the input alone. The network can also only work on a predetermined input and output size, so either scaling algorithms must be used or the network must be retrained for each input/output size. As there are unique weights for every point in the domain, translationally invariant rules must be explicitly implemented into the training data, which requires either a very large dataset or careful data augmentation, both of which significantly increase training time.

Inspired by the biology of the human visual system, convolutional layers were introduced by LeCun et al. \cite{LeCun1998Gradient-basedRecognition}, in which the weights and connections are instead expressed as a convolution between a kernel and the previous layer.
\begin{equation}
    \textbf{x}^{n} = \sigma^{n} \left(\textbf{W}^n \star \textbf{x}^{n-1}\right)
\end{equation}
Weights are now shared across the entire domain and each point in a layer is only connected to a small neighbourhood of the previous layer, allowing for much easier learning of local, translationally invariant rules, like the differential equations that govern elasticity. Further removing any fully connected layers allows for a fully convolutional neural network \cite{Long2015FullySegmentation}, which can natively be applied to any input size.

\section{Related Work}
While machine learning is a field which dates back to the 1950s, it wasn't until the last decade with the advent of Deep Learning and Convolutional Neural Networks that machine learning began to be applied to physical problems due to the now abundant memory and computional speed available. 

DeVries et al. \cite{DeVries2017EnablingAcceleration} applied a traditional fully connected network to approximate full field solutions to viscoelastic problems using only a handful of randomly distributed samples of the solution. The network takes as input the global coordinates of the point of interest and returns the displacement vector at that point, learning the function mapping between these two variables based upon a small number of samples. This allows for significant acceleration over running the full viscoelastic calculation while encoding the solution in a compact, portable form. With the weights and biases and knowledge of the network architecture, the solution can be easily reconstructed using freely available machine learning libraries rather than requiring complex simulation software. 

Fully connected networks have also been used to estimate the stress on a patient's aorta given its shape \cite{Liang2018}. The position of 5000 points on the surface of the aorta are given as input to an encoder, which generates a shape code, that is then passed through a network that converts it to a stress code that is unpacked by a decoder into the stress at each of the 5000 points on the surface. Both the encoder and decoder were trained in an unsupervised manner as part of an autoencoder, creating an efficient low dimensional representation of the shape and stress data. Not only does this significantly speed up the estimation of stress, the workflow also lends itself to further automation by using machine learning to estimate the shape of the aorta from medical imaging. This alleviates the need for manual construction of a finite element model, a time-consuming and labour intensive task. 

Other approaches to stress estimation employ convolutional neural networks to exploit the locality of the governing equations and a natural encoding of geometric features. StressNet and SCSNet proposed by Nie et al. \cite{Nie2020StressNetworks} use convolutional layers to estimate the stress in a given cantilever geometry, where StressNet is fully convolutional with some fully connected residual paths while SCSNet uses some fully connected layers in the centre of the network. Prediction error of these networks is around 10\% and 2\% for SCSNet and StressNet, respectively, while runtimes are significantly faster, taking seconds to calculate hundreds of configurations that would otherwise take hours with a traditional FE analysis approach, allowing for real-time optimisation in structural design.

Machine learning approaches have also been used to aid modelling of a variety of microstructural phenomena. A deep learning approach has been applied to estimating stress and strain fields in a two phase composite with plastic deformation by Yang et al. \cite{Yang2021DeepComposites}. In this work, a generative-adversarial network (GAN) was trained to produce the von Mises stress and plastic strain in a given composite sample following a compressive loading-unloading test. A GAN, two components are required. A generator network, in this case a variant of U-Net, is tasked with producing a predicted stress/strain field, while a discriminator network, in this case PatchGAN, is trained to detect whether a given stress/strain field is real data or generated by the generator network, with both networks trained in tandem. The composite samples consist of an 8x8 image consisting of two phases, soft and hard, to produce a 256x256 map of the stress and strain in the domain. FE analysis was used as the ground truth.

Incorporating a CNN into the physics-informed framework allows for the direct transfer from the input configuration to the output fields, so can be used as a general solver for a specific PDE. Zhao et al. \cite{Zhao2023Physics-informedData} uses a network based on U-Net \cite{Ronneberger2015U-net:Segmentation}, which takes a heat source field as input and outputs the associated temperature field, solving the heat transfer equation generally for any arbitrary source field. The hard boundary constraint method was used to include the boundary conditions as an intermediate step in the loss calculation. As the boundary conditions are not included as a network input, the network can solve arbitrary source fields but is limited to a single boundary configuration. While that limits this framework as a fully general solver, many problems such as tensile tests assume the same boundary conditions with only variation in the internal structure.

In this work a 3D grain structure, with a variety of textures, under uniaxial elastic tensile loading is used to train a U-net convolutional network.  The output of the network is a microscale stress map incorporating the grain to grain compatibility stresses. The intention of this work is to feed this map into the model of patel et al. for \cite{Patel2021AAlloys} prediction of hydride precipitation to improve its predictive ability without comprising its great advantage of speed.

\section{Methodology}
\subsection{Dataset Generation}
Grain structures were simulated using controlled Poisson Voronoi tessellation using the VGrain software package \cite{Zhang2011AnGeneration} \cite{Zhu2014TheTessellations}. A domain size of 10 μm x 10 μm x 10 μm was used with average volumetric grain size of 10 μm$^3$ and 20 μm$^3$, corresponding to a linear grain size of 2.7 μm and 3.4 μm and resulting in 100 or 50 grains per simulation respectively. The grain structures were generated with regularity, $\alpha$, of 0.73, where regularity is defined as in Zhang et al. \cite{Zhu2014TheTessellations} and parameters from Zhu et al. \cite{Zhu2014TheTessellations} due to the 3D structure.
\begin{align}
    \alpha &= A\left(\frac{c}{c_m} - \frac{c_0}{c_m}\right)^{k+nc/c_m}, \quad c_0 \leq c \leq c_m \\
    P_r\left(c\right) &= \int_{D_L/D_\mathrm{mean}}^{D_R/D_\mathrm{mean}} \frac{c^c}{\Gamma\left(c\right)} x^{c-1} \exp\left(-cx\right) dx
\end{align}
where $P_r, D_L, D_R$ and $D_{\textrm{mean}}$ are input parameters and c is implicitly calculated from equation (6). $\Gamma(x)$ is the gamma function which extends the definition of the factorial to the real numbers. $\alpha = 0$ corresponds to a pure Voronoi tessellation while $\alpha = 1$ results in the most regular structure, for example a hexagonal lattice in the 2D case.

VGrain generates Python scripts that can be imported into the ABAQUS FE analysis software for simulation of the stress response of polycrystal structures.  The built-in functionality for the material properties was used with the fully anisotropic elasticity tensor. The Python scripts were then modified to include the elastic stiffness tensor of Zr material properties at 298 K from Tromans' review of HCP elasticity \cite{Tromans2011ElasticPolycrystals}. Orientation was incorporated using ABAQUS' built-in coordinate orientation functionality, with orientations generated from a uniform distribution over $SO(3)$, the 3D rotation space. This was chosen to give the neural network the widest range of potential grain misorientations for learning the stress response without bloating the training set size, allowing for a smaller training set size and faster training. 
\begin{figure}[htp]
    \centering
    \includegraphics[width=8cm]{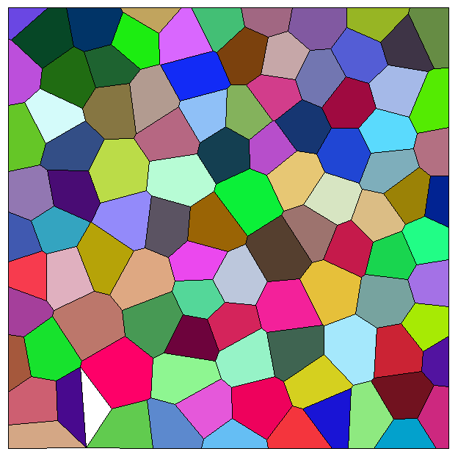}
    \caption{Example of VGRAIN structure. The colours of each grain correspond to the Euler angles in the ZXZ basis, with red, green and blue channels mapped to the respective Euler angle.}
    \label{fig:grains}
\end{figure}

Meshes in ABAQUS were generated using unstructured quadratic tetrahedral elements (C3D10) as this allowed for automatic mesh generation for all simulations without human input, which would be unfeasible for larger dataset sizes. Elements were generated using a seed spacing of 0.3 μm along grain edges, resulting in around 2 million elements per simulation. The simulations were loaded using a uniform stress of 1 MPa tensile along the positive X/Transverse face of the domain with encastre constraints along the negative x/transverse, using the linear elastic implicit solver. The chosen load is abitrary thanks to the linearity of the elastic stress equations, so the load and resulting stress field can be made dimensionless.
\begin{gather}
    C = L \hat{C} \\
    \sigma = C : \varepsilon = (L \hat{C}) : \varepsilon = L(\hat{C}:\varepsilon) = L\hat{\sigma}\\
    \nabla \cdot \sigma = \nabla \cdot (L \hat{\sigma}) = L (\nabla \cdot \hat{\sigma}) = 0 \rightarrow \nabla \cdot \hat{\sigma} = 0
\end{gather}
where $C$ is the elastic stiffness tensor, $\sigma$ is the stress, $L$ is the applied load and variables with a hat are the dimensionless counterparts.
\begin{figure}[htp]
    \centering
    \includegraphics[width=8cm]{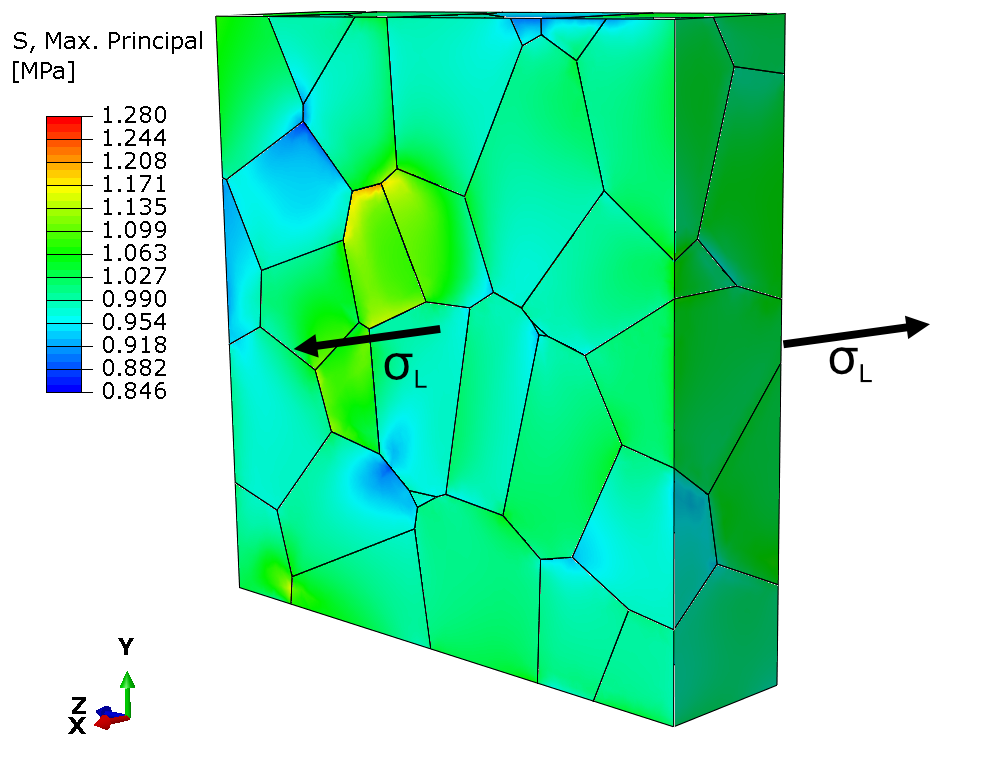}
    \caption{Example of FEA results from ABAQUS, showing the maximum principal stress of a cut through through from the model. Locations of stress concentrations and depressions can be seen at the grain boundaries of up to 28\% of the initial loading stress of 1 MPa.}
    \label{fig:FEA}
\end{figure}

The ABAQUS simulations were run on a workstation with a 2.3 GHz quad core Xeon and 64 GB of RAM, taking around 43 hours to complete with each sample taking around 13 minutes. GPGPU acceleration was not used as it caused erratic changes in the time taken to complete, so it was unclear if doing so would have benefited simulation times overall.

Once the stresses had been calculated in ABAQUS, the output database files were exported to the VTK file format using code based on the work of Liu et al.\cite{Liu2017ParaViewMicrostructures} to be voxelised in the Paraview \cite{Ayachit2015TheApplication} data visualisation software using the "Resample to Image" function. A resolution of 128$^3$ voxels was used as a compromise between resolving stress gradients and the memory footprint of the samples. The samples were then saved to the VTI format for import into Python. 200 samples were generated, with 180 used for gradient descent training and 20 used for validation during training. While this is a relatively limited dataset size for most machine learning applications, initial experimentation found much better than expected performance and therefore this size was used going forward.

An additional 165 samples were generated for validating the trained network against a variety of microstructural regularities and textures. Five distinct textures were tested, with the Kearns factors used in Table \ref{tab:Kearns}, with 33 samples per texture. The samples were split across 11 different regularities, $\alpha = 0.0$ to $0.9$ in 0.1 increments, with an additional three samples at $\alpha = 0.73$ to match the training data. None of the samples in this validation set were used in the training of the network. Non-uniform textures were generated by rejection sampling of the crystal $c$-axis using a double-poled Bingham-Kent distribution on the unit hemisphere, with the rotation matrix completed by taking perpendicular vectors, rotated by a uniformly random angle around the crystal $c$-axis.

\begin{table}[htp]
\centering
\caption{Kearns factors for the textures tested where $R$, $T$ and $N$ correspond to the $y$/rolling, $x$/transverse and $z$/normal directions respectively.}

\begin{tabular}{l|l|l|l}
\hline
\rowcolor[HTML]{C0C0C0} 
Texture               & $f_R$        & $f_T$        & $f_N$        \\ \hline
Strongly Normal     & $0.10$       & $0.10$       & $0.80$       \\
Weakly Normal       & $0.25$       & $0.25$       & $0.50$       \\
Uniform             & $0.\bar{33}$ & $0.\bar{33}$ & $0.\bar{33}$ \\
Weakly Transverse   & $0.25$       & $0.50$       & $0.25$       \\
Strongly Transverse & $0.10$       & $0.80$       & $0.10$      
\end{tabular}
\label{tab:Kearns}
\end{table}

A further 200 samples were generated for evaluating the effect of increasing the training dataset size on the performance of the network, maintaining the 9:1 ratio of training to validation samples. These samples were all generated with a uniform random texture, a regularity, $\alpha$, of 0.7 and average volumetric grain size of 10 μm$^3$, corresponding to a linear size of 2.7 μm.

Crystal orientations are encoded using a mapping in $\mathds{R}^4$ of the $c$-axis vector on the projective hemisphere, $\mathds{RP}^2$. Due to the symmetry of HCP crystal systems, rotation around the $c$-axis does not change the elastic properties therefore does not change the stress response so can be neglected. The following map was used to map the vector into $\mathds{R}^4$ so equivalence relations are reflected in the numerical representation i.e. a rotation by $\pi$ and $3\pi$ should yield the same representation and so should antipodal points. The following mapping was used:
\begin{align}
    r_1 &= x y\\
    r_2 &= y z\\
    r_3 &= z x\\
    r_4 &= z^2 - (x^2 + y^2)/4
\end{align}
where $r_i$ are the resulting vector components and $x, y$ and $z$ are the components of the $c$-axis on the unit hemisphere. This map produces a unique embedding for all points on the unit hemisphere so will not result in any orientations resolving to the same vector. This representation was found to improve the networks convergence so shortened training time compared to the axis representation or the use of Euler angles. The mapping is inspired by the projection of a Veronese surface into $\mathds{R}^4$, though the specific form of this map was chosen to make the distance between the mapping of two points similar to the angle between two points on the unit hemisphere.

Only the maximum principal stress is used in this work as this was considered the most useful parameter for cracking models. The full stress tensor may be modelled in future work using six channels in the output for each unique stress component. Prior to input into the network, stress is converted to a relative basis to ensure that values are $\sim O(1)$
\begin{equation}
    \hat{\sigma} = \frac{\sigma_{\mathrm{max}}-\sigma_L}{\sigma_L}
\end{equation}
where $\hat{\sigma}$ is the relative stress, $\sigma_\textrm{max}$ is the maximum principal stress of that voxel and $\sigma_L$ is the applied stress.

\subsection{Network Architecture}
U-Net \cite{Ronneberger2015U-net:Segmentation} was used as the basis for the neural network as it has been successfully adapted to a variety of problems beyond medical image segmentation while also being simple to implement. Several modifications have been made to allow 3D orientation maps to be fed in. The network is formed of blocks of two consecutive convolutions followed by either a max pooling or a transposed convolution for the encoding and decoding sides respectively, with channel concatenations between the encoder and decoder sides, allowing spatial information to filter forward across the network. 3D convolutions with a kernel size of 3x3x3 throughout the network with zero padding of depth 1 to ensure that a size of 128$^3$ is maintained across the network. Downsampling on the encoder side is implemented using max pooling in 2x2x2 regions, resulting in a halving of domain size along all dimensions. Upsampling on the decoder side is implemented using transposed convolutions with a kernel size of 2x2x2 and a stride of 2, resulting in a doubling of domain size along all dimensions. The network is organised into blocks of two convolutions followed by an up/downsampling operation, with concatenation of downsampling inputs and upsampling outputs to feed spatial information into later parts of the network. The network is implemented using the PyTorch deep learning library with the CUDA 10.2 backend \cite{Paszke2019PyTorch:Library}.

\begin{centering}
\begin{figure}[htp]
    \centering
    \includegraphics[width=17cm]{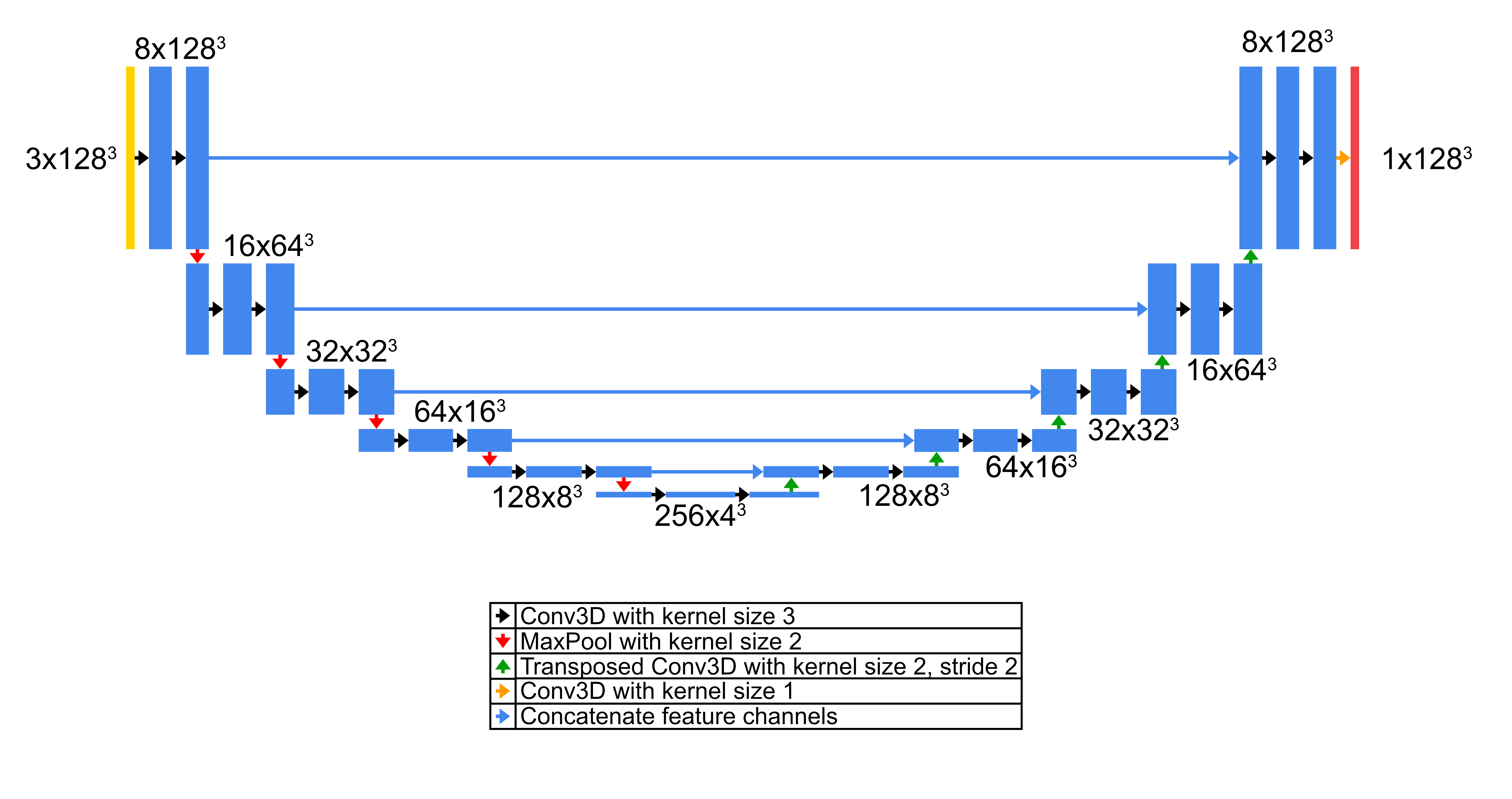}
    \caption{U-Net architecture used in this work. $N$x$M^3$ indicates a layer has $N$ feature channels and a voxel resolution of $M$.}
    \label{fig:unet}
\end{figure}
\end{centering}

An additional network architecture was also used to test the viability of reduced resolution training data. The network has the same overall architecture but the number of channels in the hidden layers has been doubled thanks to the memory freed from the smaller input and output tensors. Increasing the number of channels increases the expressibility of the network at the expense of increased memory usage and computation time and can improve convergence during training.
\subsection{Training}

The network was trained for 250 epochs with a batch size of 10 to keep the VRAM usage within 16 GB. Network weights were optimised using PyTorch's Adam optimiser, which combines standard gradient descent with a momentum term to prevent suboptimal solutions due to shallow local minima. The cost function used was a weighted sum of MSE in the absolute value of the stress and MSE in the spatial gradient of the stress, estimated using the 3D Sobel operator. The weighting between the absolute stress and the spatial gradient, $\beta$, increases from 0 for the first epoch to 0.5 for the last epoch in order to ease training of the network, since spatial gradients are very noisy at the beginning of training but become stable as training progresses. A factor $\kappa$ is tuned manually to ensure that the terms both have similar magnitudes, with $\kappa = 60$ used for this work.
\begin{align}
    C(\hat{y}|x,y) &= (1-\beta) \MSE(\hat{y},y) + \beta \MSE(\nabla_{\mathrm{Sobel}}\hat{y},\nabla_{\mathrm{Sobel}}y)/\kappa\\
    h &= [1,2,1]\\
    h' &= [1,0,-1]\\
    H_{x,ijk} &= h'_i h_j h_k\\
    H_{y,ijk} &= h_i h'_j h_k\\
    H_{z,ijk} &= h_i h_j h'_k\\
    \nabla_{\mathrm{Sobel},l} y &= H_l \star y
\end{align}
Training was performed using an nVidia Quadro RTX 5000 16 GB and took around 5 hours to complete. As the spatial gradient of the stress is important in diffusion problems, the spatial gradient was monitored during validation as well as the stress itself. 

\begin{centering}
\begin{figure}[htp]
    \centering
    \includegraphics[width=17cm]{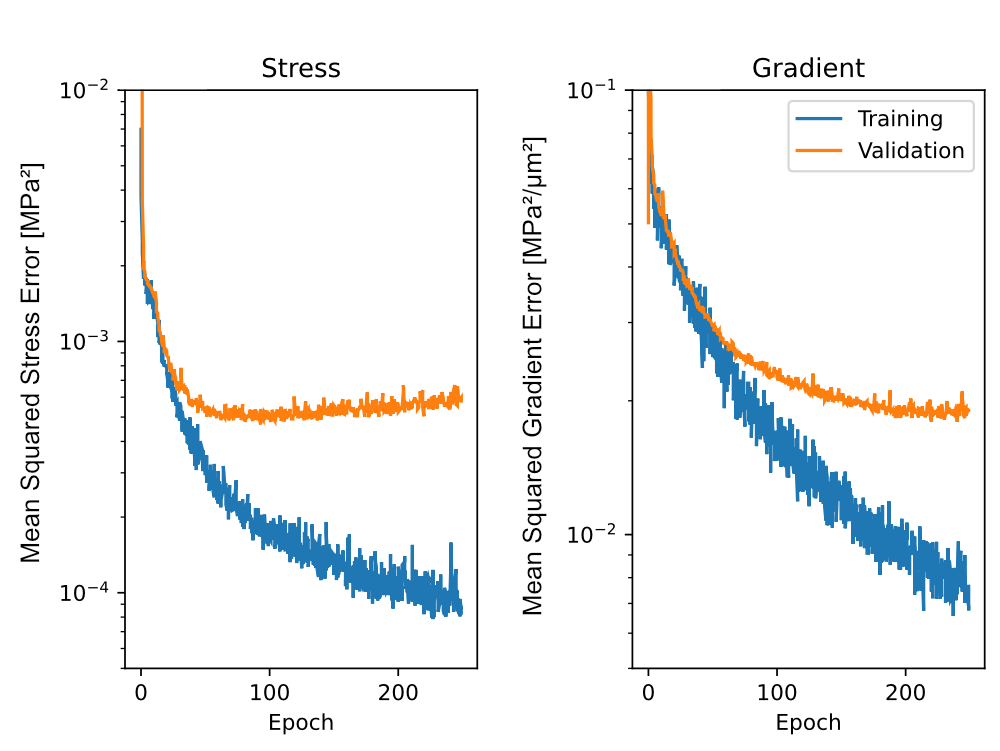}
    \caption{Mean squared error of the network output and spatial gradient of network output.}
    \label{fig:train_stats}
\end{figure}
\end{centering}

Figure \ref{fig:train_stats} shows that while error in the validation stress gradient continues to decrease until around 200 epochs, the error in the validation stress stops improving at around 75 epochs, suggesting error in the result is driven by a lack of training data as opposed to network saturation. This is unsurprising given the limited size of the training dataset so it is expected some configurations had insufficient representation in the dataset.

\subsection{Error Estimation}
Prediction errors in the output for each dataset are estimated by running the respective dataset through the trained neural network to produce the corresponding predictions. The absolute value of the difference for each sample between the prediction and the ground truth is then taken for further analysis, which will be referred to as the absolute residual. Mean errors for each dataset are calculated by taking the mean of the absolute residual for each sample while 99th percentile and maximum errors are calculated by sorting each sample into a list and taking the corresponding value. Additionally, basis errors are calculated using the assumption of the entire domain being composed of a single homogeneous crystal and finding the absolute residual.

For hypothesis testing, the mean of the means is taken then tested with the Shapiro-Wilkes test to ensure normality before a one-tailed t-test is used to quantify confidence that including the network is on average an improvement over assuming a homogeneous crystal for downstream processing.

\section{Results and Discussion}
\subsection{200 samples with 128$^3$ resolution}
\begin{centering}
\begin{figure}[htp]
    \centering
    \includegraphics[width=17cm]{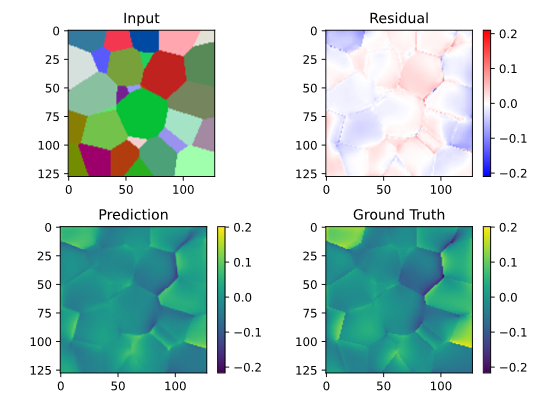}
    \caption{Example of a network prediction with; (a) representation of input orientations, (b) the residual (difference between the prediction and the ground truth), (c) the network predicted stress and (d) the FE ground truth. Stresses are in MPa.}
    \label{fig:pred_example}
\end{figure}
\end{centering}

Figure \ref{fig:pred_example} shows the output of the network and demonstrates a modest accuracy sacrifice in return, shown in Figure \ref{fig:pred_example}.d for a significant speed-up compared to a full FEA treatment. In most cases, the network correctly identifies areas of higher/lower stress than the load with similar extremes of stress. In particular the network is very effective at identifying the stress gradients in the structure, displaying significantly lower error rates compared to the neglecting grain compatibility stresses.

The largest errors in the network are driven by errors across grain boundaries, mainly due to the discontinuities between stresses across the boundaries. This causes causes the network to assign a stress according to the neighbouring grain. There are also large errors, on the order of 10\%, driven by the failure of the network to identify some grain configurations as generators of large stress deviance, likely caused by a lack of representation of the particular environment in the training set.

The bulk of the error in the output occurs due to a slight underestimate of around 2\% of the most extreme stresses, likely because these occur at the grain boundaries where the data is discontinuous, which are difficult for the network to model.

\begin{table}[htp]
\caption{Time taken for different processors to run the network on the dataset with a comparison to FEA.}
\centering
\begin{tabular}{l|r|r|r}
\hline
\rowcolor[HTML]{C0C0C0}
Processor              & \multicolumn{1}{l|}{Total Time (s)} & \multicolumn{1}{l|}{Average Time (s)} & \multicolumn{1}{l}{Speed Up} \\ \hline
Machine 1: Quadro RTX 5000        & 25                                  & 0.125                                 & 6226x                        \\
Machine 2: GeForce GTX 1050       & 77                                  & 0.385                                 & 2022x                        \\
Machine 2: i5 8300H               & 997                                 & 4.99                                  & 156x                         \\
Machine 1: Xeon Silver 4112       & 1693                                & 8.47                                  & 92x                          \\
Machine 1: FEA on Xeon (Baseline) & 155660                              & 778                                   & 1x                          
\end{tabular}
\label{tab:timings}
\end{table}
To estimate the speed-up of the network, the time taken to execute the network across the entire dataset on two machines, Machine 1 on which the network was trained and Machine 2, which was a consumer grade laptop. Table \ref{tab:timings} shows that using neural network inference is significantly faster than running a full FEA treatment, requiring around 1\% of the time on the same CPU. Introducing hardware acceleration improves time significantly further, with an order of magnitude improvement on an entry level GPU and further improvement with more powerful hardware. 

\subsubsection{Generalisation}
Performance of the network was evaluated on datasets with different grain size regularities and orientation distributions to test generalisation of the network outside the training dataset. Since the network was trained using only a single regularity and orientation distribution, it was not guaranteed that the network would be applicable to the broader space of polycrystal combinations.

\begin{figure}[htp]
    \centering
    \subfigure[]{\includegraphics[width=0.48\textwidth]{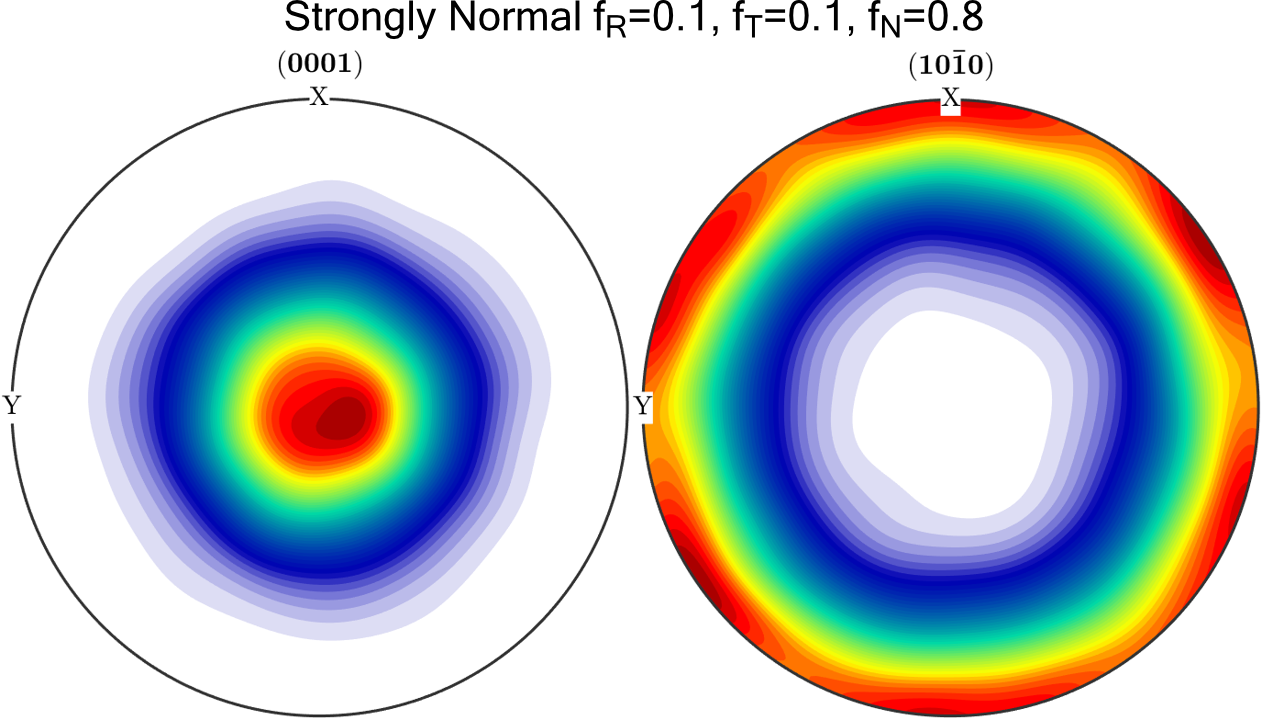}}
    \subfigure[]{\includegraphics[width=0.48\textwidth]{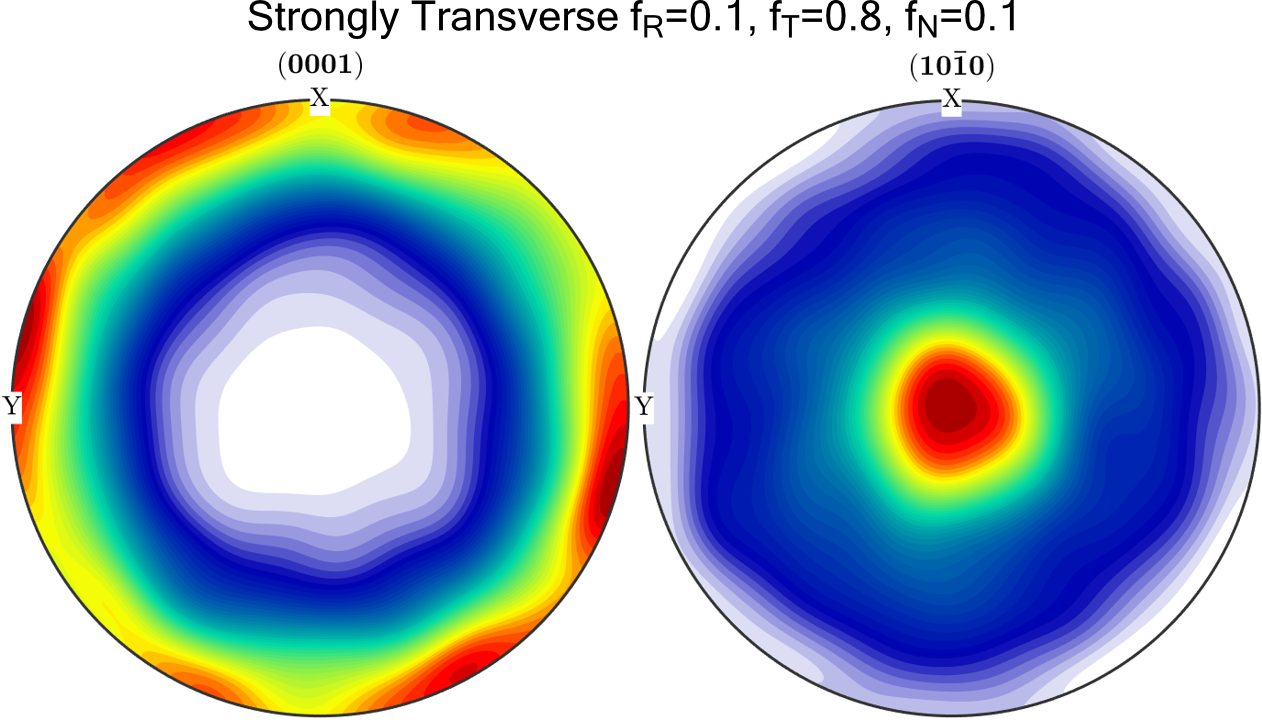}}
    \subfigure[]{\includegraphics[width=0.48\textwidth]{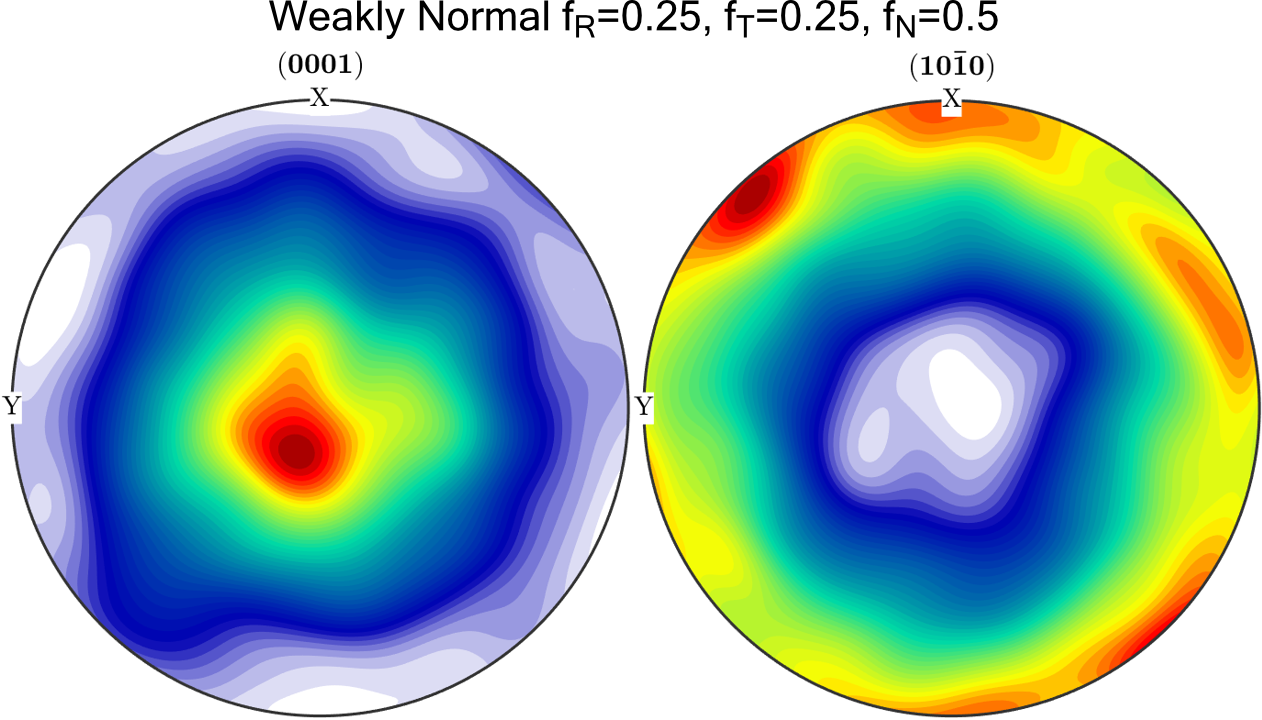}}
    \subfigure[]{\includegraphics[width=0.48\textwidth]{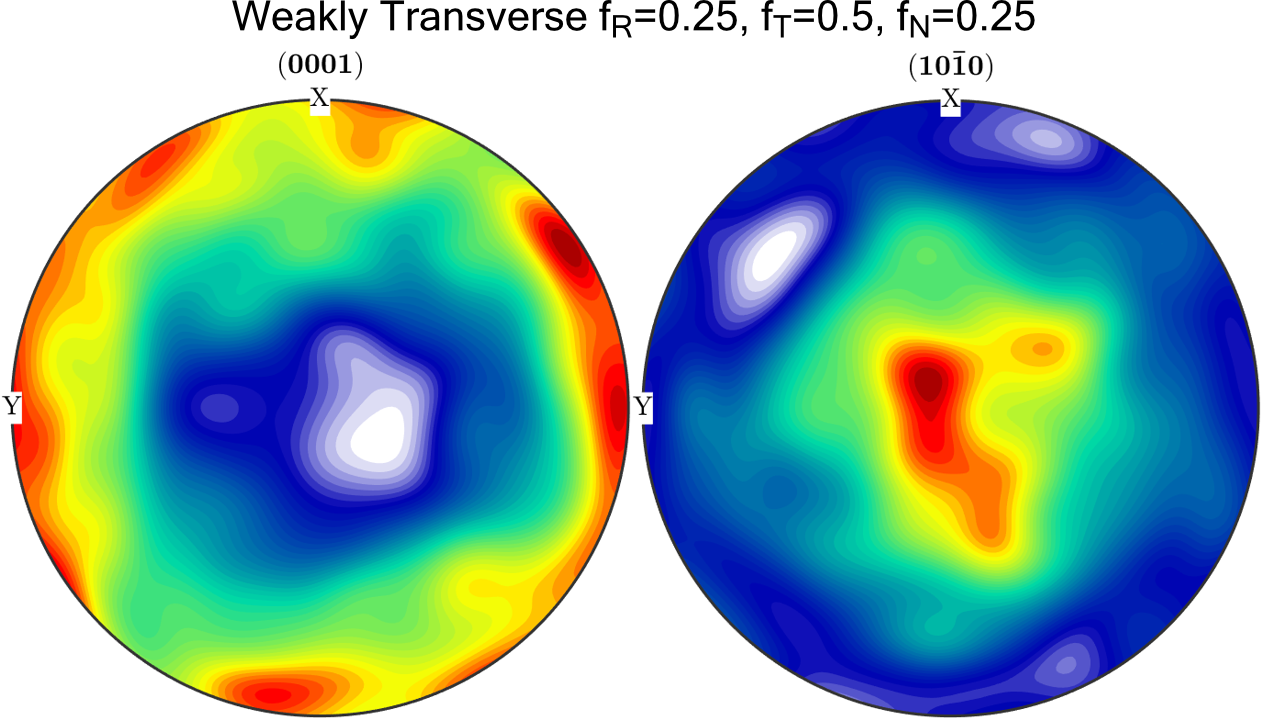}}
    \caption{Pole figures for the non-uniformly textured datasets.}
    \label{fig:pole_figures}
\end{figure}\textbf{}

\begin{figure}[htp]
    \centering
    \includegraphics[width=10cm]{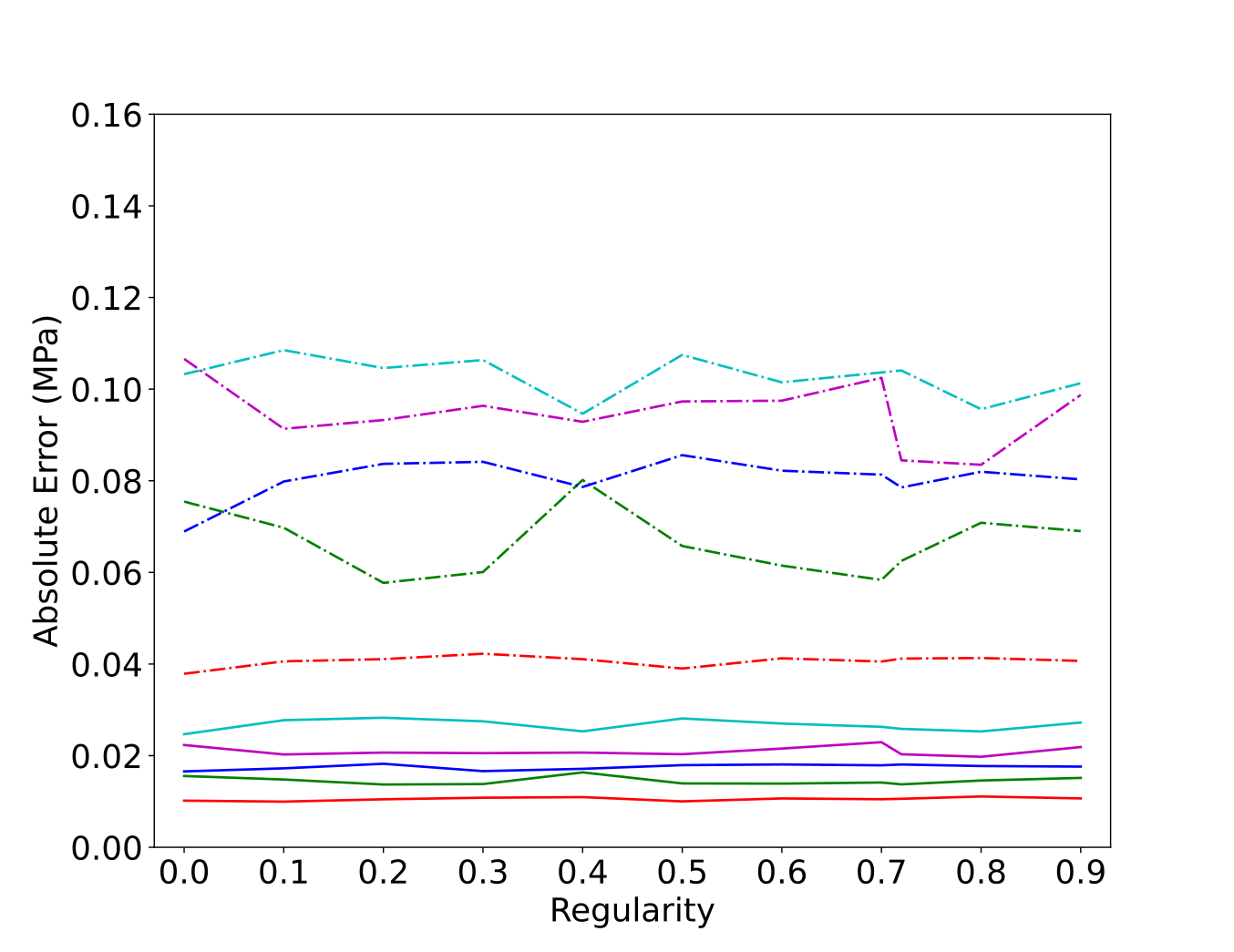}
    \caption{Graph of mean (solid) and 99th percentile (dashed) absolute error for all textures against regularity.}
    \label{fig:errors_reg}
\end{figure}

Figure \ref{fig:errors_reg} shows that the network error is insensitive to the regularity of the input grain structure for all textures, greatly expanding the range of inputs with which the network is applicable. Mean errors are essentially equal across all regularities, while 99th percentile deviation is not correlated with regularity at all, so is likely an artefact of small sample sizes. Insensitivity to the larger-scale structure of the input suggests that the network focuses on the local neighbourhood around a point to determine the stress in the output. 

\begin{figure}[htp]
    \centering
    \subfigure[]{\includegraphics[width=0.45\textwidth]{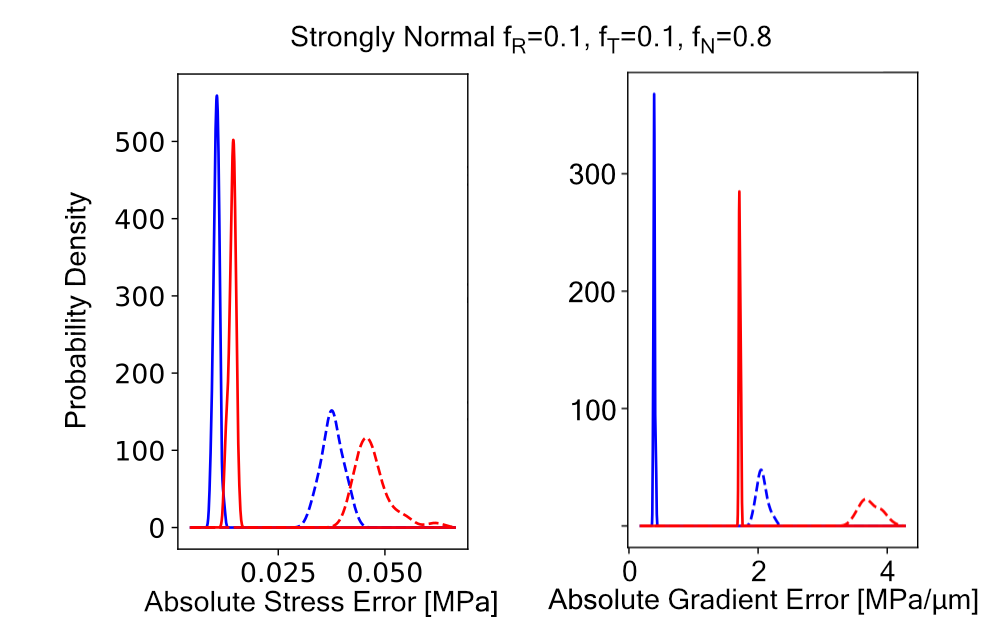}}
    \subfigure[]{\includegraphics[width=0.45\textwidth]{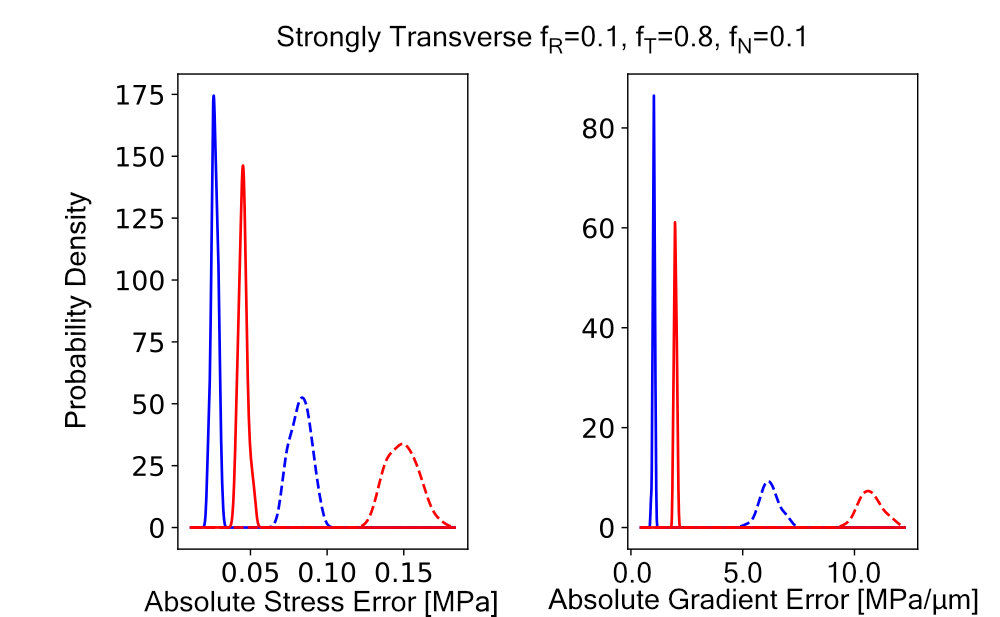}}
    \subfigure[]{\includegraphics[width=0.45\textwidth]{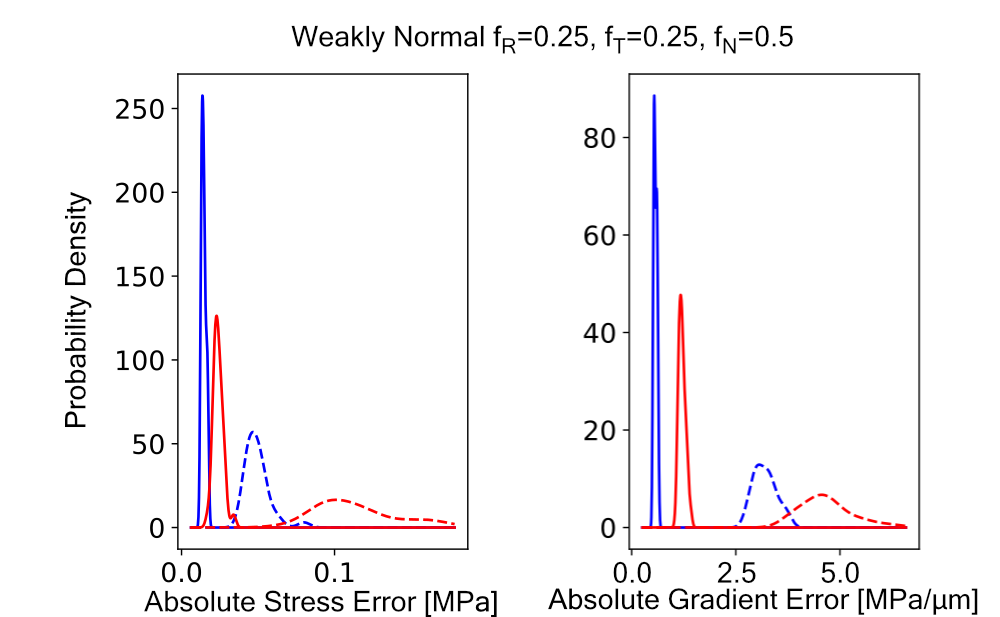}}
    \subfigure[]{\includegraphics[width=0.45\textwidth]{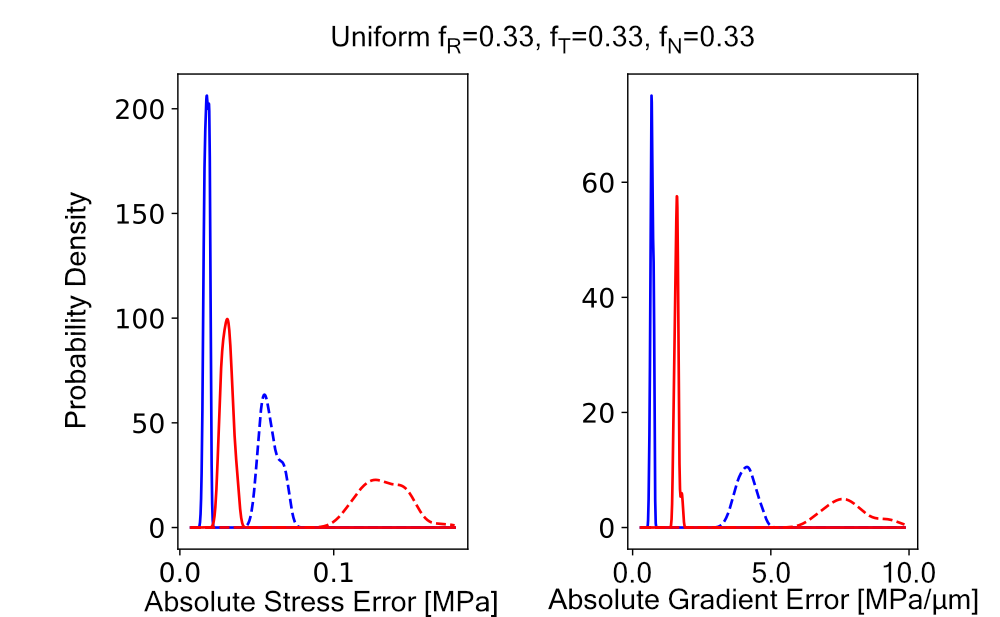}}
    \subfigure[]{\includegraphics[width=0.45\textwidth]{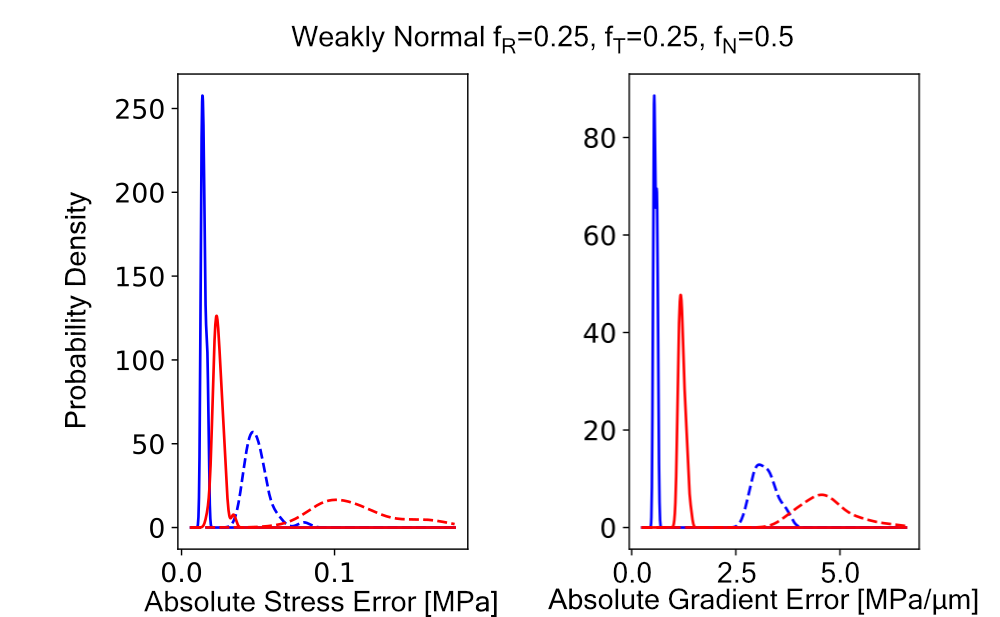}}
    \caption{Probability density distributions of mean (solid) and 99th percentile (dashed) errors of the stress and spatial gradients of the stress. The blue lines are the distributions of inference errors while red lines are the distributions of the absolute stress deviation from the loading stress.}
    \label{fig:errors_tex}
\end{figure}

There are some significant differences in absolute error, however this is reflected in significant differences in the overall stress distribution due to texture. Figure \ref{fig:errors_tex} shows how the error populations differ with respect to texture, showing significant differences between the distributions. Some of this difference can be attributed to differences in stress distribution due to texture, with mean errors less than half of the mean stress and 99 percentile errors fairing much better in all cases except for strongly transverse case. Gradient errors display a significant reduction in error, particularly in the strongly transverse ($f_R$=0.10, $f_T$=0.80, $f_N$=0.10) texture, making this network particularly useful for further calculation involving stress gradients, for example diffusion problems. 

Overall, given that the network was only trained using samples with random textures ($f_R$=0.33, $f_T$=0.33, $f_N$=0.33), comparable error distributions between the different textures suggests that the network can generalise to textures beyond that which it has been trained with. This significantly reduces the data generation needed to fully cover the input parameter space.

A one-tailed t-test analysis was performed on mean errors of each texture to ensure statistical confidence that the mean error was less than the basis case. Since the mean of the mean errors was used, the Central Limit Theorem ensures that the criterion will be close to normally distributed so a t-test is valid and this was confirmed using the Shapiro-Wilkes test at the 1\% significance level. For both the stress and the stress gradient and for all textures, the mean error is statistically significantly better than the base case, with p-values at least as small as $1\times10^{-19}$.
For the 99th percentile errors the Shapiro-Wilkes test was used to determine normality, however the distributions were not normal so t-test analysis is not valid. However, as can be seen from \ref{fig:errors_tex}, the 99th percentile errors are significantly better than the basis case for all but the the strongly transverse texture stress. For the strongly transverse case, the 99th Percentile errors are still better in the majority of samples. 

\begin{table}[htp]
\centering
\caption{T-test statistic and p-value for the mean stress and stress gradient being less then the mean of the basis case.}
\begin{tabular}{c|c|c|c|c}
\hline
\rowcolor[HTML]{C0C0C0} 
Value                             & Texture             & Kearns & Statistic & p-Value  \\ \hline
                                  & Strongly Transverse & $f_R$=0.10, $f_T$=0.80, $f_N$=0.10 & -30.9  & 1.22$\times 10^{-38}$ \\
                                  & Weakly Transverse   & $f_R$=0.25, $f_T$=0.50, $f_N$=0.25 & -25.8  & 4.36$\times 10^{-28}$ \\
                                  & Uniform             & $f_R$=0.33, $f_T$=0.33, $f_N$=0.33 & -20.7  & 1.37$\times 10^{-24}$ \\
                                  & Weakly Normal       & $f_R$=0.25, $f_T$=0.25, $f_N$=0.50 & -15.4  & 8.10$\times 10^{-20}$ \\
\multirow{-5}{*}{Stress}          & Strongly Normal     & $f_R$=0.10, $f_T$=0.10, $f_N$=0.80 & -20.8  & 6.24$\times 10^{-30}$ \\ \hline
                                  & Strongly Transverse & $f_R$=0.10, $f_T$=0.80, $f_N$=0.10 & -65.3  & 9.07$\times 10^{-58}$ \\
                                  & Weakly Transverse   & $f_R$=0.25, $f_T$=0.50, $f_N$=0.25 & -66.2  & 1.59$\times 10^{-53}$ \\
                                  & Uniform             & $f_R$=0.33, $f_T$=0.33, $f_N$=0.33 & -95.0  & 1.46$\times 10^{-67}$ \\
                                  & Weakly Normal       & $f_R$=0.25, $f_T$=0.25, $f_N$=0.50 & -26.3  & 2.77$\times 10^{-29}$ \\
\multirow{-5}{*}{Stress Gradient} & Strongly Normal     & $f_R$=0.10, $f_T$=0.10, $f_N$=0.80 & -199  & 3.68$\times 10^{-83}$
\end{tabular}
\label{tab:stat_test}
\end{table}

\subsection{Evaluation of network trained on 64$^3$ samples applied to 128$^3$ samples}
With volumetric data, reductions in resolution can yield significant memory and computation time reductions with a much smaller corresponding reduction of linear resolution. A reduction from 128$^3$ to 64$^3$ voxel resolution results in 1/8th the memory footprint while still retaining an adequate resolution of features in the samples, facilitating larger batch sizes during training. This stabilises the gradient during gradient descent and reduces the number of steps per epoch, resulting in much faster training times. Furthermore, the additional memory freed also allows for the training of larger networks than was feasible previously.

\begin{centering}
\begin{figure}[htp]
    \centering
    \includegraphics[width=17cm]{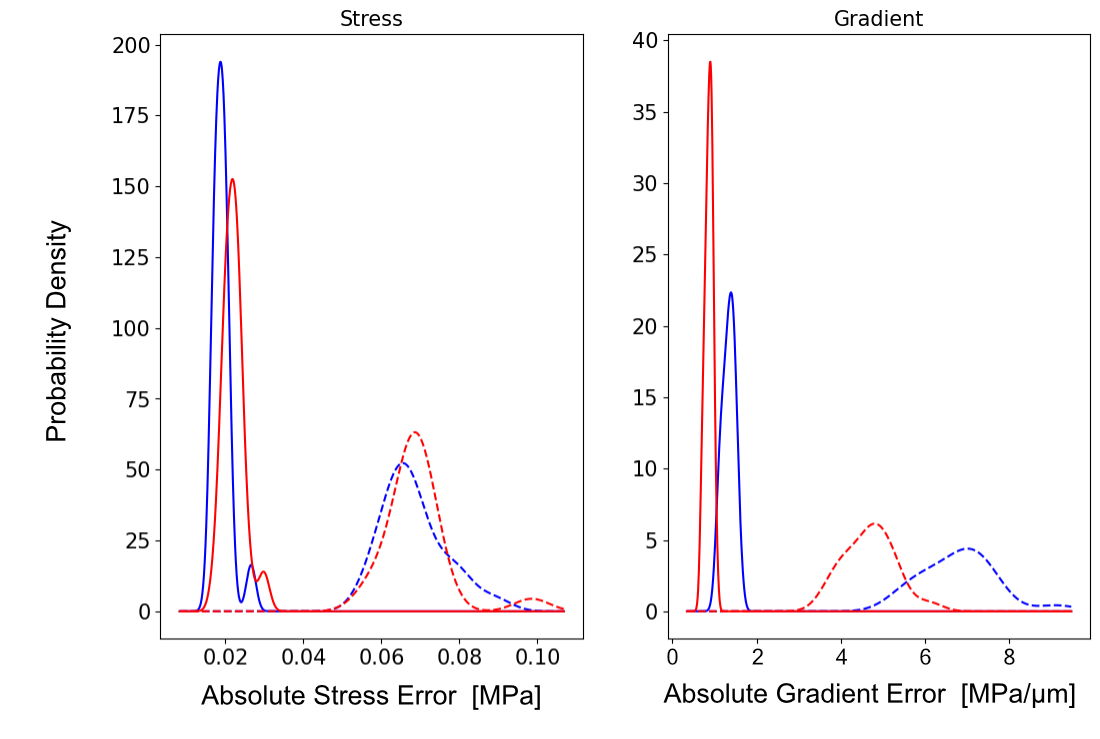}
    \caption{Probability density distribution of errors in stress and stress gradient for networks trained on $64^3$ samples applied to $64^3$ samples (blue) and $128^3$ samples (red). Mean error is solid line while 99th percentile error is dashed.}
    \label{fig:comp_res}
\end{figure}
\end{centering}

The network maintains the ability to infer the stress distribution when evaluated on higher resolution samples than it was trained with. There is some reduction in accuracy, $O(1\%)$, suggesting that there is some scaling bias, which may be addressed by including voxel coordinates in the orientation map input. On the other hand, prediction of the stress gradient improves when evaluating the network on higher resolution samples. This is possibly due to the higher resolution allowing for easier reproduction of sharp or discontinuous features at grain boundaries, reducing error in these areas.

With some modifications, scale and resolution invariance would allow the network to be applied to much larger domain sizes than would be otherwise feasible thanks to the significant reduction in memory usage and computation time. Running a full FEA on a 10 x 10 x 10 μm$^3$ domain with 100 grains typically consumes around 50 GB of RAM when using sufficient mesh density. Using FEA for larger domains would require either significantly more memory, a reduction in mesh resolution, and therefore accuracy, or paging the simulation out to storage which would considerably increase computation time. By contrast, estimation using the network requires around 200 MB of VRAM for the model itself and an additional 150 MB per 128$^3$ sample. With a 16 GB GPU, this means a domain of 40 x40 x 40 μm$^3$ with the same resolution, or an 80 x 80 x 80 μm$^3$ domain with half resolution, like in  he 64$^3$ case. This could be run entirely on the GPU, using around 9.6 GB of VRAM. Running the network on the CPU would increase inference times but would allow the network to access more memory, allowing for even larger domain sizes, possibly up to 140 x 140 x 140 μm$^3$ within the same 50 GB memory footprint of the FEA models, allowing for estimation for a domain 2744x larger than would be possible using FEA on the same hardware.

\subsection{Comparison between 200 and 400 samples at 64$^3$ resolution}
As 200 samples is a small dataset, in the context of machine learning, an additional training run including a further 200 samples was run. Doubling the size of the dataset presents a greater variety of scenarios for the network to learn on, easing training and preventing failure cases. 64$^3$ was used as the sample resolution so larger batch sizes could be used to speed up training.

\begin{centering}
\begin{figure}[htp]
    \centering
    \includegraphics[width=17cm]{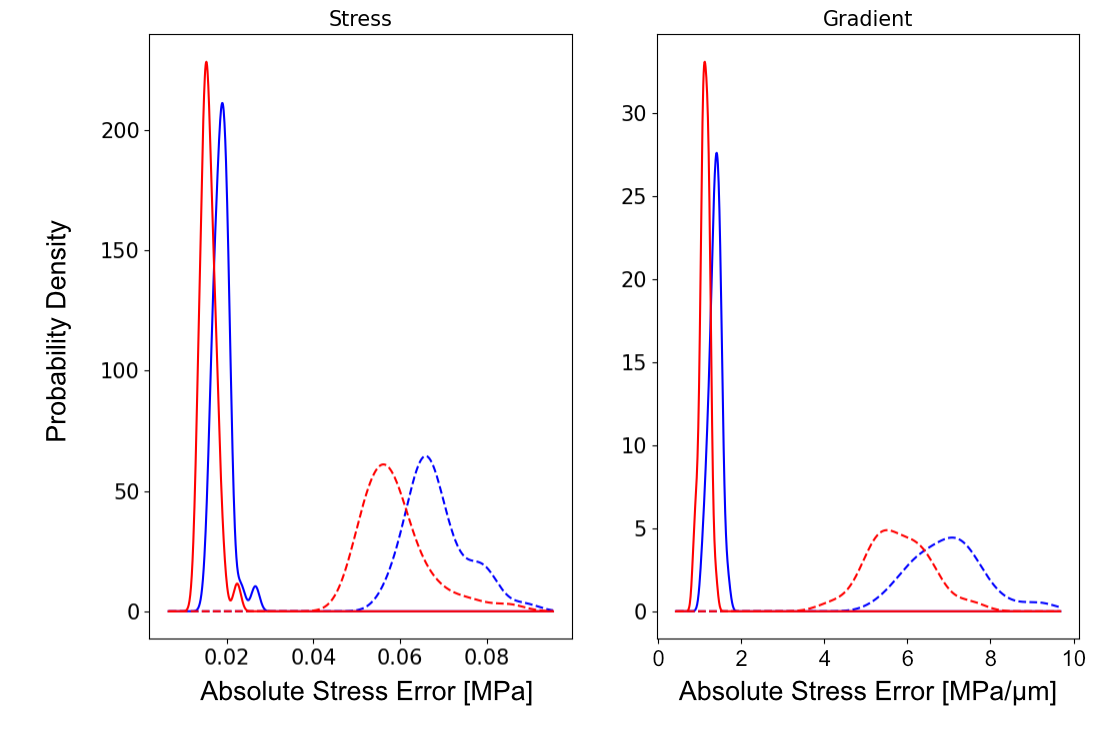}
    \caption{Probability density distribution of errors in stress and stress gradient for networks trained on 200 samples (blue) and 400 samples (red). Mean error is solid while 99th percentile error is dashed.}
    \label{fig:comp_size}
\end{figure}
\end{centering}

Table \ref{tab:error_size_comp} shows that training with the additional samples results in an increase of approximately 15\% in model accuracy compared to the 200 sample model for both the stress and stress gradient. Figure \ref{fig:comp_size} shows the mean and 99th percentile error distributions for the predicted stress and stress gradient for 200 (blue) and 400 (red) training examples, showing a consistent decrease in the errors. Additionally, the network output is much clearer with the larger dataset, representing features around grain boundaries more accurately. Maximum errors are relatively unchanged, possibly due to the maximum errors being driven by singularities in the FEA simulations. While error statistics are not significantly affected, qualitatively, the resulting estimates are much clearer with the additional training data. It's possible that this may lead to improvements in downstream simulations so further work should investigate this effect on simulation accuracy.

\begin{table}[htp]
\centering
\caption{Mean errors in stress and stress gradient for networks trained on 200 samples (ZircNet-200) and 400 samples (ZircNet-400).}
\begin{tabular}{cr|c|c|c}
\rowcolor[HTML]{C0C0C0} 
\multicolumn{2}{c|}{\cellcolor[HTML]{C0C0C0}Error Statistic} & ZircNet-200 & ZircNet-400 & \% Difference \\ \hline
 & Mean & 0.0213 & 0.0185 & -13.1\% \\ \cline{2-5} 
 & 99th Percentile & 0.0815 & 0.0706 & -13.4\% \\ \cline{2-5} 
\multirow{-3}{*}{Average Absolute Stress (MPa)} & Max & 0.2518 & 0.2552 & +1.3\% \\ \hline
 & Mean & 0.1568 & 0.1326 & -15.4\% \\ \cline{2-5} 
 & 99th Percentile & 0.7781 & 0.6735 & -13.4\% \\ \cline{2-5} 
\multirow{-3}{*}{Average Absolute Stress Gradient (MPa/μm)} & Max & 3.5420 & 3.4034 & -3.9\%
\end{tabular}
\label{tab:error_size_comp}
\end{table}

\subsection{Failure Cases}

Occasionally the network will fail to identify a grain that generates significant stress, leading to a large underestimate in the compatibility stress deviation. It is likely that these failure cases reflect a lack of representation of similar environments in the training set. Conversely the network also misidentifies a grain as a significant stress generator that does not significantly deviate. Again, this is likely due to a lack of representation in the training set. It is currently unclear which grain combinations lead to these deviations due to a lack of instances to establish a pattern.
\begin{centering}
\begin{figure}[htp]
    \centering
    \includegraphics[width=17cm]{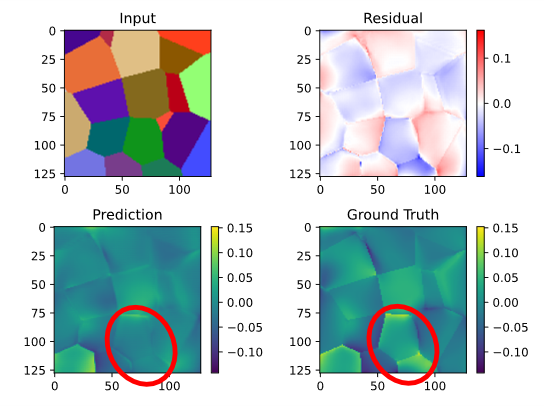}
    \caption{Example of a failure to identify a stress generating grain. The red ellipse indicates grains in which the neural network failed to predict a stress concentration that was present in the corresponding FE simulation.}
    \label{fig:fail_example}
\end{figure}
\end{centering}

Another intriguing failure to match the training data are in single points in which FEA predicts a large localised stress. These points tend to occur where the mesh elements are of low quality, with small angle tetrahedra prone to numerical instability, so it is possible that these are singularities introduced by the solver, which the network does not replicate, suggesting that a neural network approach is more resilient to numerical singularities. Additionally, in real Zr microstructures, plasticity effects mean that significant stress concentrations are unlikely to exist. This means the smoothness is positive feature for adaptation of the model to emulating plasticity simulations.

\subsection{Limitations of current model}
Currently a ``naive'' learning approach is used which does not exploit additional constraints which can be imposed on the stress field from the physics of the system. By exploiting the underlying PDEs as an additional cost term, the dependency on data could be vastly reduced compared to ``naive'' training methods, greatly expanding the generality achievable by these methods.

Additionally, the current network architecture fails to converge when attempting to train with the full stress tensor as the output. While it is possible to train an individual network for each component, this would increase computational time linearly with the number of components and may limit correlation discovery between components. The network also fails to converge when attempting to train with the displacement map as the output, limiting the effectiveness of physics informed learning as displacement boundary conditions cannot be enforced.

\section{Conclusions}
The network architecture and dataset generation combination presented is a proof of concept for a more complete system for industrial implementation. Currently, the network output is limited to the maximum principal stress, however, a full stress and strain tensor combination would be much more useful for practical problems and is the immediate focus on expansion of this work.

Generalisation of the network beyond microstructure characteristics represented in the training set has been demonstrated, greatly expanding its applicability beyond the training set. Both grain size distributions and textures beyond the training set resulted in similar inference errors, around 1\% on average, and within 10\% for 99\% of the simulation volume, when compared to those inside the training set. 

The network has demonstrated that it can capture the overall form of the microscale stress based upon only a crystal orientation map as input while requiring up to ~6000x less time and ~6500x less memory compared to traditional FEA methods, allowing for the possibility of simulating much larger domain sizes than would be possible with FE. 
%To do so would require some additional work to ensure that the network is invariant to the scale of the input. A method to achieve this that may be applied involves including spatial coordinate information as additional input \cite{Liu2018AnSolution} \cite{Facil2019Cam-CONVS:Depth} and will be investigated going forward. Eliminating any scaling bias allows the network to be applied to larger domain sizes without being specifically trained on larger domain sizes. 

This work uses a ``naive'' learning approach which does not utilise the constraining PDEs of linear elasticity. Despite the lack of explicit physics, the model is still able to capture the principal stress fields present in the microstructure with reasonable accuracy.

%Currently a ``naive'' learning approach is used which does not exploit additional constraints which can be imposed on the stress field from the physics of the system. By exploiting the underlying PDEs as an additional cost term, the dependency on data can be vastly reduced compared to ``naive'' training methods, greatly expanding the generality achievable by these methods.
\section{Acknowledgements}
This work was completed with funding from EPSRC Centre for Doctoral Training in Nuclear Energy Futures (EP/S023844/1)and Rolls-Royce Plc. (UK).
\section{Data Availability}
The processed data and models are available on request.
\bibliography{references.bib}
\end{document}